\documentclass[doublecol,showpacs,preprintnumbers,amsmath,amssymb]{epl2}
\usepackage{color,latexsym,epsfig,amssymb,amsfonts,amsmath,graphicx,bbm,enumerate,bm}

\newcommand{\ket}[1]{\left | \, #1 \right \rangle}

\newcommand{\fir}[1]{Fig.~\ref{#1}}

\begin{document}

\title{Efficient spatially-resolved multimode quantum memory}

\author{K.~Surmacz \and J.~Nunn \and F.~C.~Waldermann \and K.~C.~Lee \and Z.~Wang \and I.~A.~Walmsley \and D.~Jaksch}
\shortauthor{K.~Surmacz {\it et al.}}

\institute{Clarendon Laboratory, University of Oxford, Parks Road,
Oxford OX1 3PU, United Kingdom}

\abstract {We propose a method that enables efficient storage and
retrieval of a photonic excitation stored in an ensemble quantum
memory consisting of Lambda-type absorbers with non-zero Stokes
shift. We show that this can be used to implement a multimode
quantum memory storing multiple frequency-encoded qubits in a
single ensemble, and allowing their selective retrieval. The
read-out scheme applies to memory setups based on both
electromagnetically-induced transparency and stimulated Raman
scattering, and spatially separates the output signal field from
the control fields.}

\date{\today}

\maketitle

The distribution of entangled states over large distances is a
basic requirement for realizing quantum networks and numerous
quantum communication protocols \cite{ekert,bennett}. It is also
of importance for fundamental experimental tests of quantum theory
\cite{tittel}. As recently pointed out \cite{durlimit} current
entanglement distribution protocols using quantum repeater
stations between two parties \cite{briegeletal,duretal} will only
work over large distances if significantly improved quantum
memories become available. Ideally these memories should also be
able to store multiple qubits and allow their selective retrieval
\cite{simon}. Here we propose a scheme to achieve these goals for
photonic qubits stored in a quantum memory (QM) consisting of an
ensemble of three level $\Lambda$-type absorbers as shown in
\fir{Fig_levels_setup}(a).

Our scheme is based on the approach
\cite{fleischhauerlukin1,polzik,kraus,chaneliere,julsgaard,phillips,hau}
where the qubit is encoded in a \emph{signal} light pulse, and
stored in the ensemble via a classical \emph{control} field [see
\fir{Fig_levels_setup}(a)] propagating colinearly with the signal
in the positive $z$-direction [as defined in
\fir{Fig_levels_setup}(b)]. Storage of the signal produces a
highly-asymmetric spin wave in the medium \cite{josh,gorshkov}.
After some time the pulse can be retrieved by applying another
control field. If this field also propagates in the positive
$z$-direction, the spin wave of the medium has a low overlap with
the read-out mode, resulting in a low memory efficiency unless one
resorts to a much larger coupling strength for read-out. Instead
one can reverse the read-out control field, which maximizes this
overlap. However for non-zero Stokes shift $\omega_{13}$ [see
\fir{Fig_levels_setup}(a)] the memory efficiency suffers due to
momentum conservation. It is desirable to have $\omega_{13}\neq0$
for independent addressability of the transitions
$\ket{1}\leftrightarrow\ket{m}$ and
$\ket{3}\leftrightarrow\ket{m}$, as selection rules are rarely
stringent enough to ensure this. Furthermore, in colinear quantum
memories one also has to spectrally separate the retrieved signal
field from the stronger control field. For a weak signal this
filtering is difficult, and another method for resolving the
fields, such as orienting the control field slightly off-axis
\cite{boller,moiseev}, would be advantageous.

\begin{figure}
\begin{center}
\includegraphics[width=8cm]{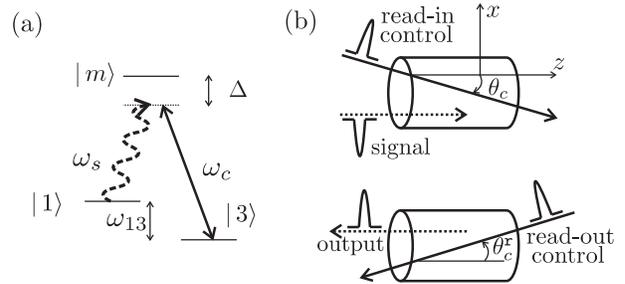}
\end{center}
\caption{Scheme for efficient storage and phasematched
spatially-resolved read-out in an ensemble QM. (a) Absorber level
configuration, and (b) schematic experimental setup. A signal
pulse (frequency $\omega_s$, wavevector $\mathbf{k}_s$) is stored
by a control field (frequency $\omega_c$, wavevector
$\mathbf{k}_c$). The configuration is such that initial state
$\ket{1}$ is energetically higher than storage state $\ket{3}$.
The control field is angled so that the resulting spin wave has
zero $z$-momentum, allowing efficient retrieval of the signal in a
different direction to the read-out control field.}
\label{Fig_levels_setup}
\end{figure}

The aim of this paper is to overcome these limitations in the
read-out. Our proposal applies to QMs based on both stimulated
Raman scattering \cite{raymer1,raymer2,josh} and
electromagnetically-induced transparency (EIT)
\cite{fleischhauerlukin1,chaneliere,phillips,hau}.  We show that
using our method, a multimode QM can be realized whereby
frequency-encoded qubits are stored in and retrieved selectively
from a single ensemble. This would allow manipulation of stored
qubits, for example in an optical lattice QM \cite{muschik},
without needing to use multiple ensembles for the different
logical states. The number of modes that can be stored in this way
is limited experimentally by angular resolution and the achievable
frequency of signal and control fields. In a conventional
single-mode memory the signal and control pulses are colinear and
the levels are chosen so that state $\ket{1}$ is energetically
lower than $\ket{3}$. In this case $\omega_{13}\neq0$, and the
stored spin wave has a momentum
$\bm{\Delta}\mathbf{k}=\mathbf{k}_s-\mathbf{k}_c$ (we call this a
\emph{phase mismatch}), rendering backwards read-out of the signal
inefficient. To overcome this (i.e.~to \emph{phasematch} the
system) we choose the level configuration shown in
Fig.~\ref{Fig_levels_setup}(a), and orient the read-in control
field at an angle of
$\theta_c=\cos^{-1}(|\mathbf{k}_s|/|\mathbf{k}_c|)$ to the signal
propagation direction [see Fig.~\ref{Fig_levels_setup}(b)]. This
eliminates the longitudinal phase mismatch and allows efficient
retrieval of the signal field. Furthermore the output signal field
direction, determined by momentum conservation, is spatially
distinct from the read-out control field. The scheme allows
high-fidelity retrieval of signal pulses with duration $T$
provided that $L\ll Tc$, and $T\omega_{13}\gg 1$, with $L$ the
length of the ensemble and $c$ the speed of light. In the
following we detail the application of this method to a general
QM, including a discussion of the constraints that arise and the
storage times that can be achieved. We then show how our scheme
could be used to implement a multimode Raman QM consisting of a
single ensemble, and derive an expression for the maximum number
of modes it is possible to store.

We consider the setup shown in \fir{Fig_levels_setup}. The
classical control field for read-in is oriented in the $x$-$z$
plane at an angle of $\theta_c$ to the signal. If $\theta_c$ is
small, the walk-off between the two fields can be neglected. The
control field envelope can then be represented at time $t$ and
longitudinal position $z$ by the slowly varying Rabi frequency
$\Omega(\tau)$, where $\tau = t-z/c$. The frequency of the
transition $\ket{1}\leftrightarrow\ket{m}$ is $\omega_{1m}$. We
introduce dimensionless coordinates $\zeta(z)\equiv z/L$ for the
longitudinal position; $\epsilon(\tau)\equiv
\omega(\tau)/\omega(T)$, with $\omega(\tau)\equiv \int_0^\tau
|\Omega(\tau')|^2\,\mathrm{d} \tau'$ the integrated Rabi
frequency, for the time; and $\bm{\rho}=(X,Y)$, where
$X=x\sqrt{\omega_{1m}/2Lc}$ (analogously for $Y$) for the
transverse position. We define the slowly-varying operators
$\alpha(\epsilon,\zeta,\bm{\rho})=\sqrt{\omega(T)/C}A(\mathbf{r},\tau)e^{-\mathrm{i}\chi(\tau,z)}/\Omega(\tau)$
for the signal field and
$\beta(\epsilon,\zeta,\bm{\rho})=\sqrt{L/C}B(\mathbf{r},\tau)e^{-\mathrm{i}\chi(\tau,z)}$
for the spin wave, where $A(\mathbf{r},\tau)$ and
$B(\mathbf{r},\tau)$ are the slowly-varying signal and spin-wave
amplitudes respectively, and $\chi(\tau,z)\equiv [\omega(\tau)
+|\kappa|^2z]/\Gamma$, which describes a Stark shift due to the
control field and a modification of the signal group velocity. We
decompose the phase mismatch as
$\bm{\Delta}\mathbf{k}=\bm{\Delta}\mathbf{k}_{\perp}+\Delta
k_z\hat{\mathbf{z}}$, with $\hat{\mathbf{z}}$ the unit vector
oriented along the $z$-axis. The coupling of the signal field to
the ensemble is $|\kappa|^2 = d\gamma/L$, with $d$ the resonant
optical depth \cite{gorshkov} and $\Gamma=\Delta-\mathrm{i}
\gamma$, where $\gamma$ accounts for dephasing and loss of the
optical polarization.

\begin{figure}
\begin{center}
\includegraphics[width=8cm]{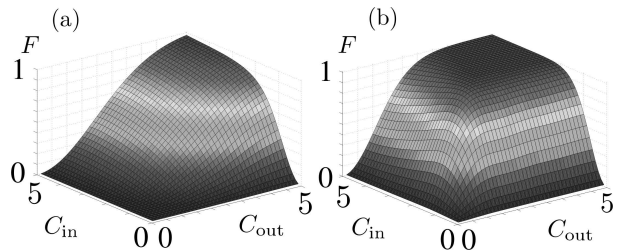}
\end{center}
\caption{Demonstration of the increase in read-out efficiency for
an ensemble QM with reverse read-out. Retrieval probability $\nu$
is plotted against the read-in and read-out couplings
$C_{\mathrm{in}}$ and $C_{\mathrm{out}}$ (defined in text). In (a)
we use colinear signal and control fields. The phase mismatch
introduced by non-zero $\omega_{13}$ degrades the efficiency, with
$\nu=0.23$ for $C_{\mathrm{in}}=C_{\mathrm{out}}=2$, which is
sufficient for storage with near-unit efficiency \cite{josh}. In
(b) the control field is angled appropriately, and
$C_{\mathrm{in}}=C_{\mathrm{out}}=2$ gives $\nu=0.95$, which is
also much higher than the efficiency achieved using colinear
read-in and colinear forward readout ($\nu=0.52$) \cite{josh}.}
\label{Fig_single_field}
\end{figure}

We assume that either $d\gamma$ or $\Delta$ are much larger than
the signal bandwidth, the maximum control Rabi frequency, and
$\gamma$, and adiabatically eliminate the excited state $\ket{m}$.
Within this limit our theory describes the interaction for
arbitrary signal and control pulse shapes in both the EIT and
Raman regimes. The absorbers are all initially optically pumped
into state $\ket{1}$, and the population of state $\ket{3}$ is
assumed to remain negligible for each absorber. In the
slowly-varying and paraxial approximations the Maxwell-Bloch
equations are
\begin{eqnarray}
\label{MB_mem_time1}
\partial_\epsilon\beta&=&-Ce^{-\mathrm{i}pX}\alpha\mbox{,}\\
\label{MB_mem_time2}
\left(\tfrac{\nabla_{\rho}^2}{4q}+\mathrm{i}\partial_\zeta\right)\alpha&=&Ce^{\mathrm{i}pX}\beta\mbox{,}
\end{eqnarray}
where $q=|\mathbf{k}_s|c/\omega_{1m}$,
$p=|\mathbf{k}_c|\sin(\theta_c)(\omega_{1m}/2Lc)^{-1/2}$, and the
coupling $C\equiv|\kappa|\sqrt{L\omega(T)}/|\Gamma|$. Note that
these equations are similar to those in \cite{josh}, except we
also include the transverse Laplacian
$\nabla_{\rho}^2=\partial_X^2+\partial_Y^2$ \cite{raymer3d}.

The dynamics of Eqs.~(\ref{MB_mem_time1}) and (\ref{MB_mem_time2})
decomposes into a linear mapping between signal and spin wave
modes. The optimally-efficient mapping excites an asymmetric spin
wave $\beta_1$ approximated by exponential decay in $\zeta$. The
symmetry of the interaction \cite{josh,gorshkov} dictates that the
optimal spin wave for forwards retrieval, $\beta_0^\mathrm{r}$, is
the mirror image in $\zeta$ of this,
i.e.~$\beta_0^\mathrm{r}(\zeta)=\beta_1(1-\zeta)$, where a
superscript $\mathrm{r}$ identifies a quantity associated with the
read-out. The efficiency of forward readout then depends on the
overlap of $\beta_1$ with $\beta_0^\mathrm{r}$, which is generally
low. Switching the propagation direction of the control field, to
readout in the backward direction, flips $\beta_0^\mathrm{r}$
around, so that its overlap with $\beta_1$ is perfect. However,
due to the residual momentum $\bm{\Delta}\mathbf{k}$ of the
spin-wave coherence after storage, we then have
$\beta^r_0\left(\zeta,\bm{\rho}\right)=\beta_1\left(\zeta,\bm{\rho}\right)\exp\left[\mathrm{i}\varphi\left(\zeta,\bm{\rho}\right)\right]$,
where
\begin{eqnarray}
\label{varphi}
\nonumber \varphi(\zeta,\bm{\rho})&=&\bigg\{\sqrt{\frac{\omega_{1m}}{2Lc}}\left(\bm{\Delta}\mathbf{k}_{\perp}+\bm{\Delta}\mathbf{k}^\mathrm{r}_{\perp}\right).\bm{\rho}+\left(\Delta k_z+\Delta k_z^\mathrm{r}\right)\\
\nonumber &&\times L\zeta
+\chi[T,L\zeta]-\chi^\mathrm{r}[0,L(1-\zeta)]\bigg\}\mbox{.}
\end{eqnarray}
The linear $\zeta$-dependence of the real terms in $\varphi$
reduces the overlap of the stored spin wave and the output mode,
leading to a reduction in the memory efficiency. This is shown in
\fir{Fig_single_field}(a) for a cesium QM, where levels $\ket{1}$
and $\ket{3}$ are the $F=3$ and $F=4$ hyperfine levels of the
$6S_{1/2}$ state respectively (i.e.~$\ket{1}$ is energetically
lower than $\ket{3}$). Our strategy is to find a configuration for
which $\Re\{\varphi\}$ is independent of $\zeta$. To do this we
choose $\ket{1}$ to be energetically higher than $\ket{3}$ and set
$\theta_c=\theta_c^{\tt{r}}=\cos^{-1}(\mathbf{k}_s/\mathbf{k}_c)$,
where the angle of the read-out control field $\theta_c^{\tt{r}}$
is as defined in Fig.~\ref{Fig_levels_setup}(b). This phasematches
the memory in the Raman and EIT configurations -- far from
resonance the $\chi$ functions in $\varphi$ are real but
negligibly small, and on resonance they are imaginary, so do not
contribute to the mismatch. Figure \ref{Fig_single_field}(b) shows
that this results in a high-fidelity read-out in $^{133}$Cs, where
now $\ket{1}$ and $\ket{3}$ are the $F=4$ and $F=3$ hyperfine
levels respectively. Note that a transverse phase mismatch does
not necessarily degrade the readout efficiency; any residual
$\bm{\Delta}\mathbf{k}_{\perp}$ simply modifies the angle at which
the retrieved signal field is emitted. However, having
$\theta_c^{\tt{r}}\neq\theta_c$ requires a modification of the
frequency of the control field to ensure that momentum and energy
are conserved.

We have assumed that the walk-off along the $z$-axis between the
signal and control fields is negligible, i.e.~small compared to
the pulse durations, so that $\theta_c\ll\sqrt{Tc/L}$. A more
stringent condition is that the transverse walk-off of the beams
is small compared to their beam waists, which requires
$\theta_c\ll\sqrt{\mathcal{A}}/L$, with $\mathcal{A}$ the
cross-sectional area of the control pulse. As well as setting the
angle $\theta_c$, the Stokes shift imposes an upper limit to the
control field bandwidth, since there should be no control photons
at the signal frequency. Diffraction limits the length of the
ensemble according to the condition $\mathcal{A} \sim \lambda L$,
where $\lambda$ is the control wavelength. These conditions can be
combined into $Tc\gg L\gg \sqrt{\mathcal{A}}$, with the maximum
storage bandwidth given by $T\omega_{13}\gg~1$. The upper bound on
$L$ is relaxed if the control field focussing is loosened, but the
control pulse energy should then be increased. However, in
general, the above considerations imply that the number density
required for high-efficiency ($>$90\%) phasematched operation of a
$\Lambda$-ensemble QM is given approximately by $n=\Theta T^{-2}$,
with $\Theta\simeq1\un{m^{-3}s^{2}}$, in both EIT and Raman
configurations.

In a $^{133}$Cs QM the $6S_{1/2}$ state is split by the hyperfine
interaction with the nuclear spin ($I=7/2$), to give the $F=3$
(our $\ket{3}$), and $F=4$ states (our $\ket{1}$), with
$\omega_{13}=9.2\un{GHz}$ . The $6P_{3/2}$ state ($D_2$ line),
which lies $351.7\un{THz}$ ($852\un{nm}$) above state $\ket{1}$,
would be used for the excited state $\ket{m}$. If
$\Delta=10\un{GHz}$, then the phasematching angle is
$\theta_c\simeq0.5^{\circ}$. This would enable high-efficiency
storage and retrieval of a single photon with $T=250\un{ps}$ in an
ensemble with $L\simeq 2\un{cm}$. This bandwidth requires a number
density of around $10^{19}\un{m^{-3}}$, which can be achieved in a
cesium vapor at a temperature of $T_e\simeq360\un{K}$. For a
thermal vapor we must consider motional dephasing of the memory,
as the phase introduced by the read-in control field will average
out \cite{molmer,lukin}. The stored spin wave oscillates at the
frequency $c|\mathbf{k}_c|\sin\theta_c\simeq
c|\mathbf{k}_c|\theta_c$, and so the entanglement fidelity of the
memory \cite{surmacz} is given by
$\mathcal{F}\simeq\exp[-|\mathbf{k}_c|^2\theta_c^2t_s^2k_BT_e/M]$,
with $t_s$ the storage time of the memory, $k_B$ Boltzmann's
constant, and $M$ the atomic mass. For the above parameters the
maximum storage time with $\mathcal{F}>0.9$ is
$t_s\simeq200\un{ns}$, i.e.~three orders of magnitude larger than
the signal pulse duration. Longer storage times could be achieved
by using a storage unit consisting of an ultra-cold gas, but this
requires a larger sample than has currently been achieved
experimentally.

\begin{figure}
\begin{center}
\includegraphics[width=8cm]{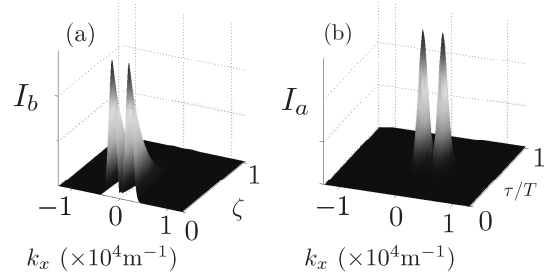}
\end{center}
\caption{Storage and retrieval of two signal modes with different
frequencies: (a) the stored spin wave intensity $I_{b}$, and (b)
the retrieved signal field intensity $I_{a}$. Both are
Fourier-transformed in the $x$-coordinate to illustrate the
different spectral components. The two signal modes require
control fields angled at $\theta_{c}=\pm3^{\circ}$, thus creating
two components to the spin wave at either side of $k_x=0$.
Read-out of the two modes is performed simultaneously with a
control field, consisting of two frequency components, angled at
$\theta_c^{\tt{r}}=3^{\circ}$. One output photon propagates along
the $z$-axis, and the other propagates at an angle of $6^{\circ}$
to the $z$-axis. Alternatively two separate control fields, one
after the other, could be used to retrieve the modes at different
times.} \label{Fig_two_phot_demo}
\end{figure}

Using the off-axis QM scheme described above, it is possible to
store multiple signal fields of different frequencies in a single
ensemble, and retrieve them selectively, as demonstrated in
\fir{Fig_two_phot_demo}. Hence one can implement a multimode QM
\cite{simon} in which frequency-encoded qubits are stored, and
retrieved either one after the other, or at the same time but
spatially separated. To see this we consider the same setup as in
\fir{Fig_levels_setup}, but now the signal field consists of two
components, labeled 1 and 2, of frequency $\omega_{s,1}$ and
$\omega_{s,2}$ respectively. Signal component $j$ ($j=1,2$) is
Raman resonant with control field $j$, which has frequency
$\omega_{c,j}$, detuning $\Delta_j$, and is angled at
$\theta_{c,j}$ to the signal for phasematching. Both signal
components can be stored in the ensemble, and the resulting spin
wave has two components separated in $k_x$ (where $k_x$ is the
momentum component in the $x$-direction), due to the different
transverse momenta of the control fields [see
\fir{Fig_two_phot_demo}(a)]. Backwards read-out of each mode can
be performed using a control pulse with the appropriate
orientation and frequency, in analogy with the single mode case.
These control fields could be applied at different times to read
out the modes one after the other, or the angles can be changed so
that the modes are emitted in different directions [as shown in
\fir{Fig_two_phot_demo}(b)]. We ensure that
$|\Delta_1-\Delta_2|\gg(\delta_1+\delta_2)$, with $\delta_j$ the
bandwidth of signal component $j$, so that signal field $j$ cannot
initially interact with control field $i$ ($i\neq j$). This
condition requires that $\max[\Delta_1,\Delta_2]$ is large,
i.e.~in the Raman regime. However, once a material excitation is
created by the absorption of signal $j$, this excitation may
interact with control field $i$.

To calculate the form of the stored spin wave and the retrieved
signal field, we again assume that the angles relative to the
signal field of both control fields are small. Control field $j$
has wavevector in the $x$-direction with rescaled magnitude $p_j$.
We assume that control field $j$ has Rabi frequency
$\Omega_j(\tau)=\Omega_jf(\tau)$, with $\Omega_j$ the maximum Rabi
frequency, $f(\tau)$ an envelope function normalized so that
$\omega_j(T)=|\Omega_j|^2T$, and the integrated Rabi frequency
$\omega_j(\tau)=\int_0^{\tau}|\Omega_j(\tau')|^2\mathrm{d}\tau'$.
Also due to the large number of absorbers in our ensemble, the
probability of a single absorber being excited by both signal
components is assumed to be negligible. The coupling constant is
now $C_m=\kappa\sqrt{LW}$, where
$W=\sum_{j=1}^2\omega_j(T)/|\Gamma_j|^2$. The dimensionless
operator for signal component $j$ is written as
$a_j(\epsilon,\zeta,\bm{\rho})=\sqrt{T/C_m}e^{-\mathrm{i}\chi_j(\tau,z)}A_j(\mathbf{r},\tau)/f(\tau)$,
and the spin wave operator is now
$b(\epsilon,\zeta,\bm{\rho})=\sqrt{L/C_m}e^{-\mathrm{i}\chi_b(\tau,z)}B(\mathbf{r},\tau)$,
where $A_j(\mathbf{r},t)$ is the slowly-varying amplitude of
signal field $j$, and the spin-wave momentum mismatch is now
$\bm{\Delta}\mathbf{k}=\hat{\mathbf{z}}\omega_{13}/c$. The complex
exponents are
$\chi_j(\tau,z)=\sum_{i=1}^2\omega_i(\tau)/\Gamma_i+|\kappa|^2z/\Gamma_j$,
and $\chi_B(\tau,z) = \sum_{i=1}^2\omega_i(\tau)/\Gamma_i^2$
respectively. The Maxwell-Bloch equations are thus
\begin{gather}
\label{MB_two_p1}
\big(\tfrac{1}{4q}\nabla_{\rho}^2+\mathrm{i}\partial_{\zeta}\big)a_j=C_mc_je^{-\mathrm{i}(p_jX+\zeta/R_j)}b\mbox{,}\\
\label{MB_two_p2}
\partial_{\epsilon}b=-C_m^*\big[e^{\mathrm{i}(p_1X+\zeta/R)}\bar{c}_1a_1+e^{\mathrm{i}(p_2X+\zeta/R_2)}\bar{c}_2a_2\big]\mbox{,}
\end{gather}
with $c_j=\sqrt{T/W}(\Omega_j/\Gamma_j)$,
$\bar{c}_j=\sqrt{T/W}(\Omega_j^*/\Gamma_j)$ and
$R_j=\Gamma_j\sqrt{W}/|\kappa|\sqrt{L}$; subscripts indicate the
signal components to which the quantities refer. The solution for
the two-component signal field, which can be achieved in the same
way as for the single-field case, is the convolution of the
integral kernels for the single-field interaction \cite{josh} with
terms that arise due to the mixing of the two components. This
mixing occurs, for example, when during storage signal component
$j$ interacts with both control fields by a second-order process,
and could potentially degrade the read-in. However, our
calculations have shown that the mixing only becomes significant
when the read-in coupling is $C_m \gtrsim 15$ -- much larger than
the coupling sufficient for signal storage with near-unit
efficiency ($C=2$). For $C_m=2$ we see that the storage process
for the two-component signal field is almost identical to the
results obtained when the two signal components are assumed to
interact independently of each other. The read-out process also
obeys Eqs.~(\ref{MB_two_p1}) and (\ref{MB_two_p2}), and a similar
lack of mixing ensures that the modes can be addressed
independently. Note that if the $c_j$'s have a relative phase of
$\pi$ between them, then the two absorption processes interfere
destructively, and the storage efficiency is low. For effective
storage of the signal components this phase must be zero, which
can be ensured either by changing the sign of the $\Delta_j$'s, or
by phase-shifting the $\Omega_j$'s.

Since one can in principle have arbitrarily-far blue-detuned
fields (avoiding any higher levels), the main limitations on the
number of modes that can be stored are experimental -- the
achievable optical frequencies and the possible angular resolution
$\Delta\theta$. The maximum control field angle allowed is given
by $\theta_{\mathrm{max}}=\sqrt{\mathcal{A}}/L$ and the number of
different control fields allowed is
$\lfloor{\theta_{\mathrm{max}}/\Delta\theta}\rfloor$. The smaller
angles considered here correspond to the fields being far
blue-detuned, which may be outside the optical regime. Taking this
into account, the number of modes that can be stored is given by
$n_m=\lfloor\{\theta_{\mathrm{max}}-[2\omega_{13}/(\omega_{1m}+\Delta_{\mathrm{max}})]^{1/2}\}/\Delta\theta\rfloor$,
with $\Delta_{\mathrm{max}}$ the largest allowed blue detuning
from $\ket{m}$. For the $^{133}$Cs QM discussed previously, taking
the angular resolution to be
$\Delta\theta=\lambda/\sqrt{\mathcal{A}}$, setting
$\mathcal{A}=10^{-7}\un{m}^{2}$, and restricting
$\Delta_{\mathrm{max}}$ so that no other transitions are excited,
this corresponds to $n_m\sim100$ modes. As with the single-mode
case the phase introduced by control field $j$ is
$\omega_{c,j}\theta_{c,j}/c=\omega_{13}/(c\theta_{c,j})$. Hence
the limitation on storage time of the memory due to dephasing, as
calculated earlier for a single mode, is given by the entanglement
fidelity of the mode that requires the smallest control field
angle for phasematching.

In summary, we have demonstrated a QM scheme using an ensemble of
three-level $\Lambda$-type absorbers that allows efficient storage
and phase-matched spatially-resolved read-out of the signal field.
We use this method to demonstrate how a multimode QM could be
implemented using a single atomic ensemble. Multimode memories
will enable the storage of multiple qubits in one memory setup,
significantly decreasing the resources required for entanglement
distribution protocols. Furthermore, storing multiple qubits in a
single ensemble opens up new possibilities for scalable
entanglement manipulation.

This work was supported by the EPSRC through the QIP IRC
(GR/S82716/01) and project EP/C51933/01, and in part by the
National Science Foundation under Grant No. NSF PHY05-51164. JN
thanks Hewlett-Packard and FCW thanks Toshiba for support. IAW was
supported in part by the European Commission under the Integrated
Project Qubit Applications (QAP) funded by the IST directorate as
Contract Number 015848.

\end{document}